# Comprehensive MHD modelling of ten successive CMEs driving a historic geomagnetic storm - The 2024 Mother’s Day event

Shirsh Lata Soni[1, 2], Anwesha Maharana[3, *], Sanchita Pal[4] and Stefaan Poedts[5,3]

[1]Department of Physics and Astronomy, University of Iowa, Iowa City, IA 52241
[2]Department of Climate and Space Science and Engineering (CLaSP), University of Michigan, USA, 48105
[3]Centre for Mathematical Plasma Astrophysics (CmPA), Dept. of Mathematics, KU Leuven, 3001 Leuven, Belgium
[4]Centre for Space Science and Technology, Indian Institute of Technology Roorkee, 247667 Uttarakhand, India
[5]Institute of Physics, University of Maria Curie-Sklodowska, Pl. M. Curie-Sklodowskiej 1, 20-031 Lublin, Poland

Email: sheersh171@gmail.com
∗ The author has an equal contribution

ABSTRACT

Interacting coronal mass ejections (CMEs) result in complex heliospheric structures that can dramatically enhance their geoeffectiveness compared to isolated events. A striking example of such complex structures is that of the Mother’s Day event, which occurred during 10-14 May 2024, leading to the strongest geomagnetic storm in decades. It was driven by at least ten interacting CMEs accompanied by flare and filament eruptions from the solar western hemisphere. We aim to understand its solar and heliospheric origins, variation in plasma and magnetic characteristics, and predict its geoeffectiveness. Furthermore, we focus on bringing to the community the challenges and limitations faced during themodelling of such a complex, yet important environment. We use multi-point remote observations of CMEs and employ different methods to estimate their speeds and geometry. These parameters drive the 3D magnetohydrodynamic simulations of CME evolution and heliospheric propagation in the framework of the EUHFORIA model. The best-performing simulation reproduced the arrival time of storm and peak intensity with a lead of 2 hours, and estimated the storm strength with about 70% accuracy. This work underscores the requirements and challenges involved in accurately modelling extreme events. The major bottleneck was to constrain the CME input parameters that dictate the accuracy of the data-driven simulations. Predicting the impacts of the complex heliospheric structures during this event required the modelling of the CME-CME interactions accurately. As a community, we need to invest more in observational infrastructure while improving the speed and accuracy of our MHD forecasting models.

## 1. INTRODUCTION

Coronal Mass Ejections (CMEs) are large-scale expulsions of magnetized plasma from the solar corona into the heliosphere. These eruptions play a fundamental role in solar-terrestrial interactions and space weather phenomena. When multiple CMEs are ejected in close temporal proximity, they may interact dynamically as they propagate through the interplanetary medium. The study of CME-CME interactions is critical for understanding their kinematic evolution, magnetic field restructuring, and geo-effectiveness, as such interactions can lead to enhanced or reduced space weather impacts depending on the interaction dynamics (Gopalswamy et al. 2001, Lugaz et al. 2012).

Nearly one-third of CMEs are complex CMEs (Vourlidas & Howard, 2013; Lugaz et al. 2017). CMEs accompanied by their associated interplanetary components (such as CME-driven shocks, sheaths, and magnetic ejecta, as exemplified by (Kilpua et al. 2017), have been identified as the leading cause of up to 90% of severe geomagnetic storms characterised by a disturbance storm time (Dst) index below -100 nT Zhang et al. (2007). Although a majority of these severe storms can be attributed to single CME events (approximately 60%), a noteworthy proportion (around 27%) can be traced back to complex signatures formed through interactions between individual CMEs and other transient phenomena, including other CMEs and regions of stream interaction (Zhang et al. 2007; Vennestrøm et al. 2016).

It is important to highlight that the probability of CME-CME interactions occurring in both the solar corona and interplanetary space is significantly elevated during periods of maximum solar activity (Rodr´ıguez G´omez et al. 2020). Research by Yashiro et al. (2004) and Robbrecht & Berghmans (2009) has documented instances in which the rate of CME occurrence can exceed 10 CMEs per day during these active phases. Several investigations have indicated that interactions between CMEs are likely to amplify the impact of individual CMEs on Earth, referred to as geoeffectiveness (e.g., Lugaz et al., 2017; Kilpua et al., 2021; Soni et al., 2023). However, the geoeffectiveness amplification due to CME-CME interactions is not yet reliably quantified. Soni et al. (2023); Scolini et al. (2020); Maharana et al. (2023) proposed that the geo-effectiveness of interacting successive CMEs is profoundly amplified, almost twice that of their individual impacts.

CME-CME interactions are governed by a range of factors, including the relative velocities, magnetic configurations, and longitudinal separations of the interacting CMEs (Maharana et al. 2023). Such interactions can result in mergers, in which two CMEs coalesce into a single, more complex structure, or deflections, in which one CME alters the trajectory of another (Lugaz et al. 2017). These interactions may also modify the magnetic and plasma properties of the CMEs, influencing their geo-effectiveness upon arrival at Earth (Shen et al., 2017; Temmer et al., 2012; Maharana et al., 2023). Observational and numerical studies have demonstrated that interactions between CMEs can lead to acceleration or deceleration, enhanced magnetic reconnection, and the formation of complex shock structures (Liu et al., 2014; Manchester et al., 2017).

Interacting CMEs can be classified into three categories based on their solar source regions: a) homologous (or quasi-homologous), b) sympathetic, and c) analogous. Homologous or quasi-homologous CMEs originate from the same active region and are frequently associated with homologous flares (Schmieder et al., 1984, Svestka, 1989). Sympathetic CMEs (flares) occur in relatively close solar angular and temporal proximity, creating optimal conditions for at least partial interaction (Moon et al., 2003). Analogous interacting CMEs originate from significantly (longitudinal and latitudinal) separated active regions (Soni et al., 2023).

The storm in May 2024, also popularly referred to as the 2024 Mother's Day event, has been modelled by several authors who have predicted the arrival time of the magnetic ejecta associated with CMEs (see e.g., Schmieder et al., 2025; Thampi et al., 2025). However, the different ejecta in this interval were not identified, and their magnetic field components were not modelled. In this study, we investigate the dynamics of sympathetic interactions among 10 CMEs using magnetised CME models within an MHD framework. Although MHD modelling of CMEs is common these days, the data-driven modelling of ten interacting CMEs with a special focus on predicting the magnetic field profiles is one of its kind. The primary objective of this case study is to evaluate the EUHFORIA model's performance in handling extremely complex events. In contrast, the previous validation efforts have focused on complex scenarios involving two or three CMEs (e.g., Scolini et al., 2020; Maharana et al., 2023; Soni et al., 2024); this study extends that analysis to a more intricate system.

We introduce several modifications to the model inputs and assess their impacts on forecasting accuracy. Particular attention is given to ensuring that these adjustments are based on observational data and remain suitable for operational applications. By examining the kinematic and magnetic characteristics of the CMEs, we aim to enhance our understanding of the physical mechanisms governing CME-CME interactions and their relevance to space weather forecasting. The structure of the paper is as follows: Section 2 provides a detailed description of the CMEs that led to the severe geomagnetic impact, drawing from available coronagraph, EUV observations, and in situ observations. The corresponding methodology used to determine CME kinematics and shock propagation is presented in its subsections. The MHD modelling results are presented in Section 3. In Section 4, we provide the results and discussion. Finally, Section 5 summarises the findings and provides an outlook on the broader implications of this study on CME forecasting.

## 2. EVENT OVERVIEW

Between May 8 and 11, 2024, a sequence of ten CMEs (CME1-CME10) erupted in rapid succession from the solar western hemisphere, predominantly from NOAA AR 13664, with additional contributions from a filament eruption (CME4) and NOAA AR 13668 (CME10). The rapid succession of these 10 CMEs, with overlapping trajectories and varying kinematic properties, led to complex interplanetary interactions. The merging and compression of these CMEs in transit significantly enhanced the event's geoeffectiveness, ultimately driving one of the most intense geomagnetic storms recorded in the past two decades, with a Dst value of -420 nT.

### 2.1. Constraining CME geometric parameters

To determine the kinematics and geometry of CMEs in the solar corona, we perform a 3D reconstruction of these events from two vantage points, namely SOHO and STEREO-A, utilising the graduated cylindrical shell (GCS) model developed by Thernisien et al. (2009). GCS reproduces the morphology of CMEs in the corona and allows for quick estimates of the geometric and kinematic properties of CMEs from one or more viewpoints. The outcomes obtained from the GCS fitting process serve as crucial geometrical input to our subsequent CME modelling within the EUHFORIA framework. The snapshots of the GCS reconstruction of the CMEs are provided in Fig. A1. The parameters derived through GCS reconstruction are provided in Table 1.

Table 1. Graduated Cylindrical Shell (GCS) model constrained parameters of the Coronal Mass Ejections (CMEs)- half angle (α), leading edge height (h), aspect ratio (κ), latitude (θ), longitude (ϕ), and tilt (γ). The date and time of the first appearance of CME in LASCO C2 coronagraph, source location, and the associated flare class are listed for each CME. The final column,
Past work provides the correspondence of the CMEs in this study with Pal et al. (2025) (see Table 1), who have compiled the list based on Khuntia et al. (2025); Liu et al. (2024).

| CME | Date/Time (mm/dd hh:mm) [First appearance in LASCO C2] | Source Location | $\alpha$ (deg) | $h$ ($R_\odot$) | $\kappa$ | $\theta$ (deg) | $\phi$ (deg) | $\gamma$ (deg) | Flare Class | Active Region |
|---|---|---|---|---|---|---|---|---|---|---|
| CME1 | 05/08 4:34 | S17°W09° | 29 | 9.23 | 0.32 | -20 | 15 | -33.75 | X1.0 | 13664 |
| CME2 | 05/08 6:45 | S17°W10° | 29 | 8 | 0.32 | -20 | 14 | -42 | X1.0 | 13664 |
| CME3 | 05/08 12:48 | S16°W18° | 40 | 7.25 | 0.87 | -18 | 5 | -26.73 | M8.7 | 13664 |
| CME4 | 05/08 19:24 | N16°E41° | 46 | 7.83 | 0.18 | 11 | 310 | 87.86 | Filament | 13667 |
| CME5 | 05/08 22:36 | S18°W23° | 29 | 7 | 0.40 | -13 | 15 | -30 | X1.0 | 13664 |
| CME6 | 05/08 22:36 | S19°W28° | 16 | 4.6 | 0.15 | -15 | 38 | -83 | M9.8 | 13664 |
| CME7 | 05/09 09:24 | S15°W28° | 26 | 7.62 | 0.53 | -8 | 18 | -35 | X2.2 | 13664 |
| CME8 | 05/09 17:44 | S14°W34° | 20 | 510 | 0.56 | -8 | 30 | -5 | X1.1 | 13664 |
| CME9 | 05/10 07:12 | S15°W44° | 28 | 11.8 | 0.62 | -11 | 25.4 | 58 | X4 | 13664 |
| CME10 | 05/11 01:36 | S23°W80° | 38 | 11.6 | 0.92 | -15.2 | 33.8 | 63 | X5.8 | 13668 |

Our GCS reconstructions of the 8-11 May 2024 CME sequence show several points of convergence with the geometrical parameters reported in previous works (Khuntia et al., 2025) (PW1) and (Liu et al., 2024) (PW2), indicating a largely consistent morphological evolution across independent studies.

Both PW1 and PW2 have identified only one eruption before May 8, 12.00. However, in this study, we have identified two distinct CME signatures and propose the existence of CME1 and CME2. CME3 is identified by both studies, and CME4 is consistent with the filament eruption noted by PW1. CME5 and CME7 are consistent with the identification by both studies. PW1 suggests the eruption of another CME at the same time and from the same active region as CME5. We add it to our list as CME6. CME8, CME9, and CME10 are not considered in previous studies when analysing this storm.

Similar to these works, the early eruptions (CME1-CME3) originated from a narrow longitudinal channel near the central meridian and exhibited moderate half-angles (≈25-40) and low aspect ratios (κ ≈0.8), characteristic of compact, slowly expanding flux-rope structures. The tilt angles remain predominantly negative for the majority of the sequence, but exhibit a systematic trend toward less negative values and an eventual sign reversal for the final two CMEs. This behaviour is also reproduced in our fits and supports the scenario proposed by Liu et al. (2024), in which successive eruptions progressively modify and rotate the overlying coronal magnetic arcades.

An exception in this sequence was CME4, which erupted from the northeastern quadrant. According to our 3D reconstruction, it propagated away from the Sun-Earth line (to the east). Its orientation and pronounced tilt suggest that it followed a distinct trajectory, reducing the likelihood of interaction with the eruptions directed toward the western limb. However, some studies have inferred a more Earth-directed component for CME4 (Khuntia et al. 2025), underscoring the sensitivity of GCS solutions to observer geometry and to the identification of the leading edge. In contrast to CME4, the remaining CMEs originated from a confined longitudinal sector and showed greater diversity in their propagation directions, angular widths, and expansion characteristics. Our estimates for CME4 and CME9 show longitude offsets 10-20◦ larger than those reported by Khuntia et al. (2025), suggesting stronger non-radial deflection and more pronounced channelling by the ambient streamer belt (Mostl et al., 2015). We also find that CME7 has a larger apex height and a smaller half-angle than in previous analyses (Khuntia et al., 2025), indicating reduced lateral expansion and collimated geometry (Pant et al., 2021).

Taken together, the similarities in overall morphology and the differences in specific geometric parameters highlight both the robustness and the inherent limitations of the GCS method in a rapidly evolving, multi-CME environment. The close temporal and spatial spacing of most eruptions likely facilitated compression, deflection, and partial merging during propagation, producing a complex compound interplanetary structure, consistent with the multi-CME interaction scenario emphasised in earlier studies.

### 2.2. CME Magnetic Parameters

The handedness of the magnetic flux rope, i.e., the chirality (sign of helicity), can be inferred from different morphological features (e.g., D´emoulin, 2008), Palmerio et al., 2017). In our study, images in the extreme-ultraviolet SDO/AIA filter at 171˚A (Fig. A2) suggest that AR 13664 was characterised by a positive chirality as indicated by the presence of a forward S-shaped sigmoid in the southern hemisphere.

To estimate the orientation of the flux ropes at the Sun, we use proxies based on the orientation of post-eruptive arcades (PEAs) and polarity inversion lines (PILs) (M¨ostl et al., 2008, Palmerio et al., 2017, Palmerio et al., 2018). As shown in Figure A2, the PEA formed after the eruption of CME1 was confined to the southern portion of the AR/PIL and exhibited an approximately north-south orientation. For CME2 and CME3, we observe PEAs developing along the whole PIL structure. Although a global direction from south-east to north-west can be identified, the shape of such PEAs appears to be bent in a reverse-S shape. This reflects the complexity of the underlying PIL system, making it extremely difficult to determine an unambiguous tilt of the flux ropes from such observations. Similar conclusions about the initial flux rope tilts can be obtained by considering the locations of coronal dimmings and flare ribbons (Fig. A2). Combining the GCS-derived tilt angles with the flux-rope chirality and the underlying photospheric magnetic polarity distribution inferred from HMI magnetograms, following the classification scheme of Bothmer & Schwenn (1998) and the remote-sensing inference framework of Palmerio et al. (2018), we infer an ENW-type intrinsic flux rope for CME1 and intermediate ENW-NWS flux-rope configurations for CME2 and CME3 in the low corona. CME4 exhibits a nearly north-south oriented axis and is therefore classified as a high-inclination flux rope. CME5 and CME7 are consistent with SWN-type configurations, while CME6 also exhibits a high-inclination geometry. CME8 is best described as a SEN-type flux rope, whereas CME9 and CME10 correspond to NES-type configurations, indicating an eastward-directed axial field and a progressive rotation of the erupting magnetic system during the later phase of the CME sequence.

For the EUHFORIA simulations (Section 3), we consider the tilt parameter from GCS reconstruction, which constrains the flux rope orientation in the upper corona (~5 R☉ to ~20 R☉). The derived tilts suggest that the axial magnetic field of CME1 and CME2 are close to the solar equatorial plane, while that of CME3 has an inclination of ~40 degrees with respect to the solar equator. Using the GCS tilt estimates as given in Table 1 and the solar surface observations, and assuming the flux rope chirality is conserved during Sun-Earth propagation, CMEs 1-10, except CME4, are NWS type. CME3 is associated with an intermediate ENW-NWS flux rope. We note, however, that the $\gamma_{CME}$ parameter is associated with the highest uncertainties (Thernisien et al., 2009), and it is known to be very sensitive to the subjectivity involved in performing the GCS fitting (see, e.g., Figure 5 in Shen et al., 2018), for an alternative fitting of the CMEs using the GCS model, leading to a quite different interpretation of the tilt angles in the corona). For this reason, we consider GCS reconstruction particularly uncertain.

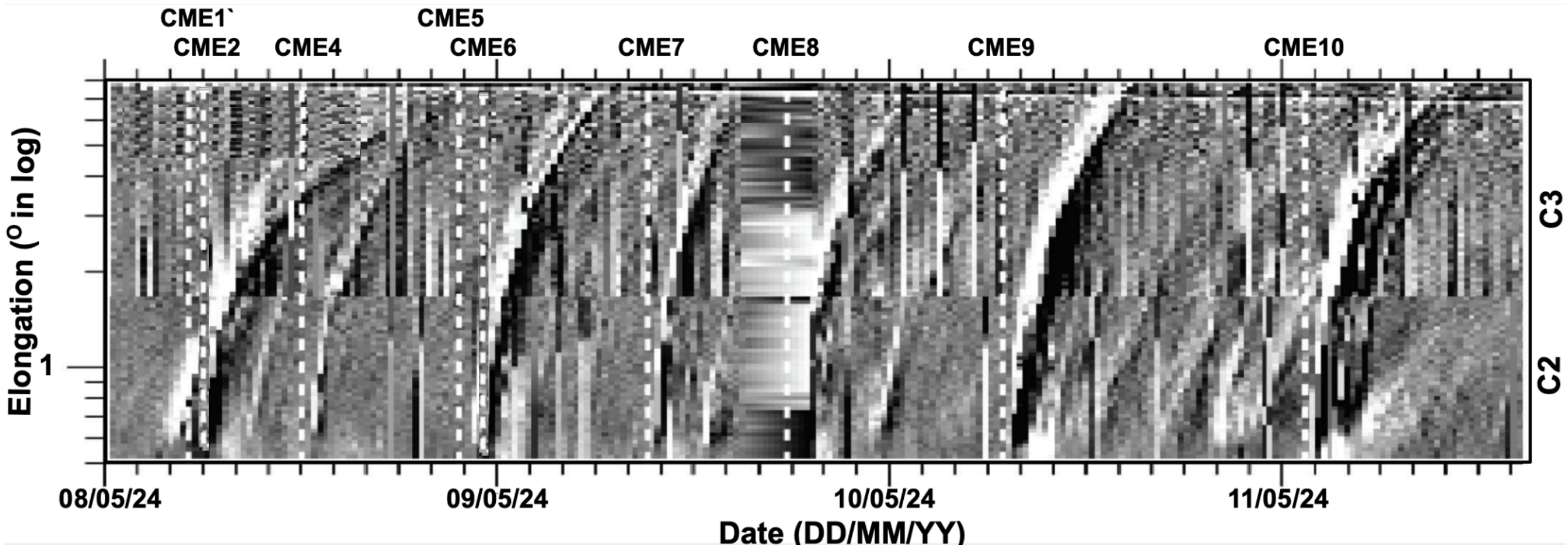

Figure 1. SOHO time-elongation map (J-map) constructed from running-difference white-light observations of the LASCO C2 and C3 coronagraphs, showing the heliospheric propagation of ten successive CMEs (CME1 to CME10) during 8–14 May 2024. The plot shows the elongation angle (degrees) relative to the Sun–spacecraft line as a function of time. Each CME appears as a distinct bright, curved track whose slope corresponds to its apparent propagation speed and whose curvature reflects Thomson-scattering geometry and radial expansion.

The magnetic field parameters are derived using the procedure followed in Maharana et al., (2024) and Scolini et al., (2020). The poloidal magnetic flux is obtained by averaging the values derived using its empirical relationship with the peak flare intensity (Kazachenko et al., 2017; Gopalswamy et al., 2017; Dissauer et al., 2018; Tschernitz et al., 2018) and the speed (Gopalswamy et al., 2017). The handedness of the EUV determines chirality. The CMEs (except

CME4) in this event erupted from AR 13664, which had a positive chirality (right-handed sigmoid) throughout the period. CME4erupted from a negative-chirality filament in the northern hemisphere, and its flux was derived using the flux-speed empirical relation.

2.3. Determination of CME Speed

Estimating the propagation speed of these CMEs was particularly challenging due to their successive ejections, interactions, and complex kinematic evolution. Given the dynamic nature of CME-CME interactions and the limitations of a single observational approach, we employed multiple complementary techniques to obtain more accurate estimates of their speeds.

First, we analysed the height-time (H-T) profiles of the CMEs using coronagraph observations from the Large Angle and Spectrometric Coronagraph (LASCO) onboard SOHO (Yashiro et al., 2004). By tracking the leading edges of each CME across different time frames in LASCO C2 and C3 fields of view, we obtained initial velocity estimates. However, due to projection effects and line-of-sight ambiguities, these measurements alone were insufficient to characterise the actual three-dimensional motion of the CMEs.

Table 2. The 3D projected speed for the ten CMEs constrained using different methods: Projected speed (from the LASCO catalogue), GCS reconstruction, J-map.

| CME | CME1 | CME2 | CME3 | CME4 | CME5 | CME6 | CME7 | CME8 | CME9 | CME10 |
|---|---|---|---|---|---|---|---|---|---|---|
| Date/Time (mm/dd hh:mm) [First appeared in LASCO C2] | 05/08 4:34 | 05/08 6:45 | 05/08 12:48 | 05/08 19:24 | 05/08 22:36 | 05/08 22:36 | 05/09 09:24 | 05/09 17:44 | 05/10 07:12 | 05/11 01:36 |
| Projected speed (LASCO) | - | 530 | 677 | 457 | 952 | 952 | 1280 | 1024 | 953 | 1614 |
| GCS speed | 265 | 397 | 378 | 450 | 462 | 1154 | 1064 | 483 | 1120 | 1238 |
| J-map speed | 428 | 513 | - | 310 | 934 | 934 | 812 | 517 | 650 | 742 |

To extend the analysis beyond the coronagraph field of view, we employed J-map tracking derived from time-elongation maps using heliospheric imaging data (see Fig. 1). J-maps, constructed from the Solar Terrestrial Relations Observatory (STEREO) Heliospheric Imagers (HI) (Howard et al., 2008), allowed us to track the CMEs as they propagated through interplanetary space. By analysing the elongation-time evolution of CME fronts, we derived183 interplanetary speed estimates that more effectively capture their acceleration, deceleration, and interaction dynamics.

Additionally, we applied the GCS model to reconstruct the three-dimensional morphology of each CME and derive a more accurate kinematic profile. The GCS fitting, using multi-viewpoint observations from SOHO/LASCO and STEREO-A, provided a robust estimation of the actual speed and angular extent of each CME. This method was beneficial for disentangling projection effects and determining the orientation of CMEs relative to the observer.

By combining these different approaches-height-time profiling, J-map analysis, and GCS modelling- we obtained a comprehensive understanding of the speeds of these CMEs. Despite being lower than the speeds adopted in previous studies (Thampi et al., 2025; Liu et al., 2024), our kinematic estimates derived from J-map and GCS tracking yield the best agreement with the observed heliospheric response when used as input to the MHD simulations at 1 AU, thereby lending strong support to the reliability of our speed determination. This multi-method approach was essential for resolving the complexities introduced by CME-CME interactions, in which individual ejecta experienced rapid acceleration, deceleration, and deflections due to preceding structures. The interplay of these dynamic factors ultimately influenced their transit time to Earth and their geo-effectiveness, contributing to the extreme geomagnetic storm that followed.

## 2.4. In-situ observation

The in situ observations during the interval of extreme geomagnetic disturbances driven by the CMEs mentioned above show signatures of complex magnetised plasma. The period May 10-12 corresponds to the most extreme geomagnetic disturbance, with Dst dropping below −450 nT, and the period May 12-14, when the Dst recovery also saw values below −100 nT. This poses challenges for distinguishing the merged from the undisturbed ejecta and for determining how many CMEs contributed to a particular merged ejecta. In this section, we will analyse the in situ solar wind observations during this period, and address the questions: (1) What is the type of the ejecta (merged/unmerged); (2) Which CME erupted between May 8-11 could have been associated with the in situ magnetic ejecta (single ICME or merged CMEs).

### IDENTIFICATION OF EJECTA

The measured data by MAG and SWE instruments onboard Wind spacecraft located at L1 is provided in Fig. 2. The interplanetary shocks during this interval are identified as S1 that reaches on May 10, and S2 on May 12 (as per Liu et al., 2024). The shaded regions in the figure were identified as the potential ejecta during that interval. Further details and the potential association of individual CMEs or merged ejecta are presented in Table 3. First, we analysed the periods of low plasma beta (β) associated with high speed (v), high total magnetic field (|B|), and low proton temperature profiles ($T_p < T_{ex}$, where Tex is the expected ambient temperature), low proton number density ($n_p$), high alpha to proton number density ($n_a/n_p$), and low non-radial velocity components (i.e., $v_y < 50$ and $v_z < 50$) to identify five regions of magnetic ejecta (namely E1, E2, E3, E4, & E5), and illustrated as shaded zones in Fig. 2). The first two ejecta, E1 (blue) and E2 (green), in this work, are the two main blocks of complex ejecta. A low |δB|/B within E3 (cyan) indicates the presence of an unperturbed CME. We note that E1 and E2 are made of multiple CMEs as pointed out by Pal et al. (2025), and E3 is also consistent with this study. The high non-radial velocity components (i.e., $v_y > 50$ and $v_z > 50$), between the shaded regions (E3 & E4; E4 & E5), indicate the presence of compressed solar wind. The ejecta E4 and E5 are intriguing. Although they are associated with high speed, β < 1 and $T_p < T_{ex}$, they lack large rotating magnetic field signatures, suggesting flank encounters.

CME-ICME CORRELATION

In Section 2, we identified ten CMEs that could have a potential geomagnetic impact on Earth during the May 10-14 period. In this section, based on in situ observations, we posit a correlation between the ten CMEs and the five complex ejecta. Pal et al. (2025) applied normalised fluctuation amplitude $|\delta B|/B$ along with the trace power spectral density (PSD) of the magnetic field fluctuations ($P_B$ ) to identify the interval of unperturbed ejecta within the larger merged ejecta. $|\delta B|/B$ is computed, where $|\delta B| = |B(t) - B(t + \tau )|$, $\tau$ = 94 seconds corresponding to the inertial range of fluctuations (Kilpua et al. 2021). The low $P_B$ and low $|B|/B$ coincided with bidirectional PAD and other characteristics of magnetic ejecta, which enabled the identification of the remains of unperturbed CMEs, and hence shed light on the number of CMEs that could have been parts of the merged ejecta E1 and E2. Using this approach, two cores were identified in E1 and three in E2.

Hence, CME1 and CME2 could be associated with E1, given a 2-hour difference in their time of appearance in the coronagraph field of view (see Tables 1 & 2). Given its higher speed, CME2 would have caught up and merged with CME1. As CME3 is slower than the subsequent CMEs, the flanks of CME4 could have accelerated it. CME5 and CME6 could have created E2. With a speed exceeding that of all six previous CMEs, CME7 (~11 hours after CME6) could have merged with E2 at its leading edge, creating the initial signatures in E3. The merged signatures in E3 could be the trailing part of CME7 interacting with CME8. As CME8 traverses even to the west of CME7, only its flank could have merged with CME7, so we do not observe a separate signature for CME7. The single isolated CME signature in E4 could be associated with CME9 as it erupted around 14 hours after CME8. CME10 erupted ~18 hours after CME9 and could be associated with E5. In this work, we evaluate these hypotheses using numerical simulations. In addition, we are studying three extra CMEs to understand the geomagnetic conditions after the main storm. We start with this preliminary hypothesis that we will explore in detail in Section 3.

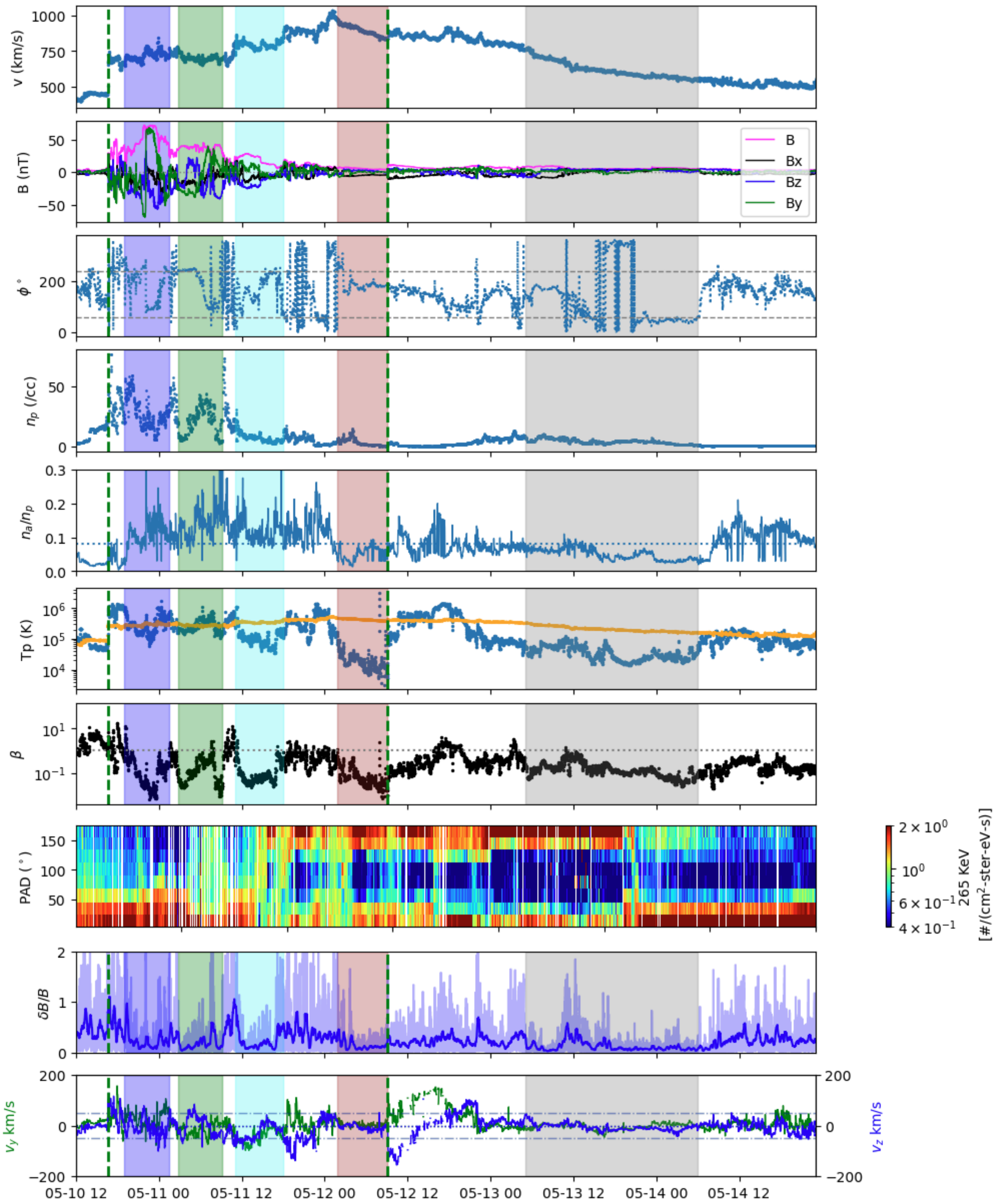

Figure 2. In situ observations using MAG and SWE instruments onboard the Wind satellite at L1. The period of the main storm, between May 10-12, and the extended enhancement, between May 12-14, are shown. The shocks S1 and S2 are marked in green dashed vertical lines. The magnetic ejecta periods are shaded as E1 (blue), E2 (green), E3 (cyan), E4 (brown), and E5

(grey). The plot shows from top to bottom: the speed v, the magnetic field components $B_x$, $B_y$, $B_z$, the total magnetic field strength |B|, the interplanetary magnetic field angle Φ, the proton number density $n_p$, Helium ion (alpha) abundance per unit density (na/np), proton temperature ($T_p$), plasma beta (β), suprathermal electron pitch angle distribution (PAD), normalised fluctuation amplitude |δB|/B, and transverse velocity components ($v_y$ , $v_z$ ).

Table 3. Identification of interplanetary CME signatures from in situ data. The colours associated with the ejecta in this table are the same as in Figure 2.

| Characteristic | Start | End | Associated CME (s) |
|---|---|---|---|
| Shock 1 | 2024-05-10 16:39:00 | - | - |
| Ejecta 1 (blue) | 2024-05-10 19:00:00 | 2024-05-11 01:30:00 | CMEs 1+2+3(+4) |
| Ejecta 2 (green) | 2024-05-11 02:48:00 | 2024-05-11 09:10:00 | CMEs 5+6 |
| Ejecta 3 (cyan) | 2024-05-11 11:00:00 | 2024-05-11 18:00:00 | CMEs 7+8 |
| Ejecta 4 (brown) | 2024-05-12 01:50:00 | 2024-05-12 09:00:00 | CME9 |
| Shock 2 | 2024-05-12 17:30:00 | - | - |
| Ejecta 5 (grey) | 2024-05-13 05:00:00 | 2024-05-13 10:40:00 | CME10 |

## 3. MHD MODELLING USING EUHFORIA245

To understand how these successive CMEs evolved after their eruption and arrived at Earth, we perform MHD simulations to model the propagation of the CMEs in the heliosphere beyond 0.1 AU (21.5 $R_\odot$). We also want to understand how the complex magnetic ejecta formed, leading to the severe geo-effective storm. We employ the physics-based space weather forecasting model EUropean Heliospheric FORecasting Information Asset (Pomoell & Poedts, 2018) EUHFORIA, to model the ambient solar wind and the CME evolution.

EUHFORIA (version 2.0) consists of a coronal model and an MHD heliospheric model. The coronal model used for this work is a 3D semi-empirical modified Wang-Sheeley-Arge model (Sheeley, 2017). In an EUHFORIA simulation, the ambient solar wind before CME onset is modelled to ensure appropriate plasma and magnetic field conditions for the CMEs to traverse. This model, driven by photospheric magnetic field observations from the synoptic magnetogram of the Global Oscillation Network Group Data Assimilative Photospheric Flux Transport (GONG-ADAPT) on May 7, 2024, at 00:00:00, computes the plasma and magnetic field conditions at 0.1 AU.

The output of the coronal model serves as the boundary condition for the heliospheric model at 0.1 AU, which numerically solves the ideal MHD equations in 3D. The resolution of the computational mesh used in this work is 0.0037 AU (corresponding to 0.78Rs) with 512 equidistant cells in the radial direction (from 0.1 to 2AU), the angular resolution is 2in the latitudinal (extending between ±80) and 2in the longitudinal (extending between 0-360) directions, respectively. CMEs are injected as time-dependent boundary conditions at 0.1 AU into the heliospheric domain. Previously, EUHFORIA was used to model this event with the non-magnetised cone model to estimate CME arrival times (Schmieder et al. 2025).

In this work, we use the linear force-free magnetised spheromak model (Chandrasekhar & Kendall, 1957) implemented in EUHFORIA and widely used in the community (Verbeke et al., 2019) to quantify the geomagnetic impact using the predicted magnetic field information. Although many advanced CME models have been developed within the EUHFORIA framework (Maharana et al., 2022; Linan et al., 2024; Maharana et al., 2024), they are currently being validated for modelling multiple CMEs. The spheromak model, due to its simple structure and fewer CME model parameters, is used in this work as a first attempt to develop a first-hand understanding of this event. In the future, the modelling of this event and geo-effectiveness predictions will be improved using the advanced CME models. The parameters required to run the spheromak model are constrained by the observed geometrical and magnetic field properties of CMEs, as mentioned in the previous section. Because the spheromak model does not consist of the CME legs, the ejecta formed by the interactions of the CME flanks could be missed. Hence, we have shifted the CMEs latitudinally towards the Sun-Earth line.

The Disturbance storm index (Dst, in nT) is a widely used measure of geomagnetic storm intensity; the sharp negative excursion in Dst reflects a strong geomagnetic storm response. It is also predicted for each simulation result using the empirical relations of (O'Brien & McPherron, 2000). We first compute Dst using the Wind measurements as input, which serves as a reference for comparing Dst predictions from our simulations (see e.g., Palmerio et al., 2023; Scolini et al., 2020; Maharana et al., 2024).

The accuracy of EUHFORIA predictions depends significantly on the observationally constrained parameters. In this study, there is a high level of uncertainty in the CME parameters due to the halo nature of CMEs, which makes it difficult to reconstruct their morphology, especially with observations from a single vantage point. We performed simulations with speeds from three sources: (1) LASCO catalogue; (2) GCS reconstruction; and (3) J-map reconstruction. The LASCO speeds are at the higher end of the estimated range, whereas GCS speeds are at the lower end (except for CMEs 6, 8, and 9). The J-map speeds are similar to the LASCO speeds for CMEs 1, 2, 4, and 5. There is no clear trend to explain the choice of speed source for CMEs. Hence, we first design CME runs with J-map speeds for all CMEs, and then present arguments (based on uncertainties) for the speed adjustments needed to reproduce the observed arrival times and geo-effectiveness data.

## 4. RESULTS AND DISCUSSION

In this section, the EUHFORIA simulations and their predictions at 1 AU are presented. Along the way, we also present the story (hypothesis) that CME parameters could be adjusted (within observational uncertainty bounds) to improve prediction accuracy, thereby highlighting potential uncertainties.

### 4.1. Real-time space weather forecaster perspective

We tested three different sets of CME speeds obtained from separate sources (see Table 2). All other kinematic parameters, including geometric and morphological properties, were taken from the GCS reconstruction and kept the same for all three cases. For CME3, the speed could not be determined from the J-map, because it was quite faint in the white-light observation. Therefore, we perform two runs: one using the 3-D speed measured by the GCS method
and the other using the projected speed for CME3. The runs are named as follows: (1) vGCS (all CMEs are launched with speeds reconstructed using the GCS model), (2) vLASCO (all CMEs are launched with speeds from the LASCO catalogue), (3) vJmap (all CMEs are launched with speeds measured using J-maps, CME3 GCS), and (4) vJmap′ (all CMEs are launched with speeds constrained using J-maps, CME3 LASCO). The arrival time (S1) prediction obtained from the four runs falls in a ~9-hour window. Without any parameter optimisation, the run ‘vJmap (CME3 GCS) performs the best in predicting the CME arrival (with 2 hours lead time). This is the best a real-time space-weather forecaster could have done if they chose this combination of speeds to obtain the best-predicted Dst, −315 nT (around 67% of the observed). If the vLASCO run (projected CME speed estimates) was relied upon, an approximate 7-8 hours delay in prediction could happen, and the dip in the Dst profile could be delayed by ~8 hours as well.

### 4.2. Post-event analysis perspective

In this section, we design the following EUHFORIA runs as experiments to determine the CME parameters with associated uncertainties. The time series of the best EUHFORIA run at Earth is provided in Fig. 3. The radial velocity ($v_r$ ) and $B_z$ in the equatorial, meridional, and transversal slices of EUHFORIA’s simulation domain, plotted in Fig. 4, illustrate the stages of evolution and interaction between the ten CMEs. In this section, we present results from sequentially launching CMEs to quantify each CME’s role in Dst prediction. The sequential simulation predictions at Earth, plotted in Fig. A3, are described as follows:

1. CME 1-2: Only CME1 and CME2 are injected. CME2, with an eruption time within 2 hours of CME1 and a higher speed, already merges with CME1 at the time of their injection into the heliospheric domain. This merged ejecta is part of E1, arriving with the shock, S1. These two CMEs alone could have resulted in a Dst~-160 nT.

Table 4. The spheromak CME model parameters of the ten CMEs that erupted in May 8-11, 2024, were used to initialise the EUHFORIA run, ‘vJmap (CME3 GCS)’.

| CME parameters | | | | | | | | | | |
|---|---|---|---|---|---|---|---|---|---|---|
| Parameters ↓ | CME1 | CME2 | CME3 | CME4 | CME5 | CME6 | CME7 | CME8 | CME9 | CME10 |
| Insertion date | May 8 | May 8 | May 8 | May 8 | May 9 | May 9 | May 9 | May 9 | May 10 | May 11 |
| Insertion time | 10:07 | 11:50 | 20:05 | 21:04 | 01:36 | 02:04 | 12:42 | 22:02 | 09:59 | 04:11 |
| Radial speed [km $s^{-1}$] | 324 | 389 | 202 | 840 | 667 | 959 | 531 | 331 | 401 | 386 |
| Latitude [°] | −10 | −10 | −8 | 1 | −3 | -5 | 0 | 0 | −1 | −5 |
| Longitude [°] | 15 | 14 | 5 | −27 | 15 | 38 | 18 | 30 | 25 | 34 |
| Radius [$R_\odot$] | 10.4 | 10.4 | 13.8 | 15.5 | 10.4 | 6.0 | 9.4 | 7.3 | 10.1 | 13.2 |
| Density [kg $m^{-3}$] | $1 \cdot 10^{-17}$ | $1 \cdot 10^{-17}$ | $1 \cdot 10^{-17}$ | $1 \cdot 10^{-17}$ | $1 \cdot 10^{-17}$ | $1 \cdot 10^{-17}$ | $1 \cdot 10^{-17}$ | $1 \cdot 10^{-17}$ | $1 \cdot 10^{-17}$ | $1 \cdot 10^{-17}$ |
| Temperature [K] | $0.8 \cdot 10^{6}$ | $0.8 \cdot 10^{6}$ | $0.8 \cdot 10^{6}$ | $0.8 \cdot 10^{6}$ | $0.8 \cdot 10^{6}$ | $0.8 \cdot 10^{6}$ | $0.8 \cdot 10^{6}$ | $0.8 \cdot 10^{6}$ | $0.8 \cdot 10^{6}$ | $0.8 \cdot 10^{6}$ |
| Helicity | +1 | +1 | +1 | −1 | +1 | +1 | +1 | +1 | +1 | +1 |
| Tilt [°] | 124 | 132 | 117 | 2 | 120 | 170 | 125 | 95 | 32 | 27 |
| Toroidal magnetic flux [Wb] | $5.1 \cdot 10^{13}$ | $5.1 \cdot 10^{13}$ | $4.7 \cdot 10^{13}$ | $4.8 \cdot 10^{13}$ | $5.1 \cdot 10^{13}$ | $5.06 \cdot 10^{13}$ | $8.0 \cdot 10^{13}$ | $1.1 \cdot 10^{14}$ | $1.4 \cdot 10^{14}$ | $4.0 \cdot 10^{13}$ |

2. CME 1-3: CME3 is added to CME 1-2. As the CME3 speed could not be obtained from the J-map method, we used the GCS speed instead, as it had a better arrival estimate. The addition of CME3 does not cause any significant change in the Dst magnitude; however, it produces two distinct delayed southward $B_z$ dips.

3. CME 1-4: CME4 is included in the CME 1-3 run. As per our GCS reconstruction, CME4 travelled 50 degrees east of the Sun-Earth line with a speed of 450 $kms^{-1}$. Using these parameters, there was no significant interaction between CME4 and CMEs 1, 2, and 3. However, Khuntia et al. (2025) estimated CME4 to be travelling 27 degree east of the Sun-Earth line at a speed 991 $kms^{-1}$. Incorporating parameters constrained by them, we found a significant interaction of the CME4 flank with the CME3. CME3 was accelerated by the overtaking CME4 to the point that it began interacting with E1 at 0.1 AU. The arrival time of E1 was shifted to only 2 hours later than its
observed arrival time. The addition of CME4 resulted in a minimum Dst that is 51% of the reference prediction and an earlier prediction of the minimum $B_z$ occurrence, which better resembles the observations than in the CME 1-3 run.

4. CME 1-5: CME5 is added to the CME 1-4 run. It begins interacting with the trailing part of CME3 and facilitates the merging of CME3 with E1, thereby further enhancing the southward $B_z$. CME5 corresponds to E2, and its trailing part extends temporally beyond E3 in the absence of successive CMEs. The inclusion of CME5 resulted in the prediction of 60% of the observed minimum $B_z$ . The arrival of the negative $B_z$ dips corresponds well with the observations.

5. CME 1-6: CME6 and CME5 have the same injection time and hence, emerge into the heliosphere already merged at 0.1 AU. With increasing momentum, E2 (combined CME5 and CME6) propagates faster, further compressing E1. 78% of the observed $B_z$ in E1 is obtained, resulting in a minimum Dst −354 nT. As CME6 has a higher speed than CME5, it dictates a westward propagation direction, and hence, the Bz in E2 decreases relative to the CME 1-5 run. Fig. 4(a) and (b) show the radial velocity, vr, and scaled Bclt (equivalent to −Bz on the equatorial plane) at a time snap during the formation of E2 before its arrival at Earth.

6. CME 1-7: Adding CME7 to the CME 1-6 run, we reproduce the speed better around May 11, 12:00. The modelled arrival time of CME7 is earlier than observed; however, the speed estimate corresponds to the beginning of E3. CME7 travels westward, but its flank starts to expand in the low-density region created by CME5 and CME6, and merges with the trailing part of CME5 around 0.6 AU. This results in the formation of the southward Bz in E3. In the absence of any successive CMEs, CME7 seems to expand excessively, and the modelled ejecta, E3, arrives late as compared to reality. In addition, the high-speed profile beyond May 11, 12:00 is not reproduced. Hence, we add CMEs 7-9, which are additional to those in previous studies (e.g., Pal et al. 2025).

7. CME 1-8: CME8 is added to the CME 1-7 run. CME8 is launched 30 degrees west of the Sun-Earth line with a speed slower than CME7. Nevertheless, its eastern flank accelerates in the low-density region ahead of it and merges with CME7, and the merged ejecta accelerates further. It is indeed the combined CME6 and CME7 that result in the formation of E3 (the observed minimum Bz is −25 nT). The modelled minimum southward Bz intensity in E3 is almost doubled in the CME 1-8 (−29 nT) run as compared to the CME 1-7 run (−12 nT) due to the CME-CME interaction. The speed associated with E3 is reproduced more accurately. The Dst time profile is in agreement with the measured trend.

8. CME 1-9: CME9 is added to the CME 1-8 run. CME9 makes a flank encounter as it is launched ~35 degrees east of the Sun-Earth line. Even if it was launched slower than CME7 and CME8, its flank accelerates in the low-density region behind E3 to hit Earth with a southward $B_z$ in its flank. This creates E4, whose southward $B_z$ is overestimated in our modelling. This could also be attributed to uncertainty in the magnetic flux estimation or to our inability to capture the exact physical dynamics in our modelling. However, the speed profile beyond E3 matches the observations significantly better. The propagation of CME9 compresses the trailing part of E3 and creates a sheath region that corresponds to the enhanced speed after E3. The modelled plasma beta and decreasing slope of speed in E4 match the observations well, confirming the presence of an expanding magnetic cloud (CME in this case). Fig. 4(c) and (d) illustrate the modelled $v_r$ , and scaled Bclt (equivalent to $-B_z$ on the equatorial plane) at a time snap during the formation of E3 and E4 in the heliosphere, and the arrival of E1 at Earth.

9. CME 1-10: CME10 is added to the CME 1-9 run. This is the last CME in this interval with a glancing blow at Earth, which creates E5. CME10 does not impact the evolution of the previous nine CMEs. However, its presence is important for reproducing the speed and plasma beta profile in E5, indicating the presence of an expanding magnetic cloud. This run suggests that even a flank encounter with a slow CME could have such high speed due to the pre-conditioning of the heliosphere by previous CME(s). Fig. 4(e) and (f) illustrate the modelled $v_r$ , and scaled Bclt (equivalent to $-B_z$ on the equatorial plane) at a time snap when all ten CMEs are propagating in the heliosphere, the formation of E5 has been initiated, and E4 reaches Earth.

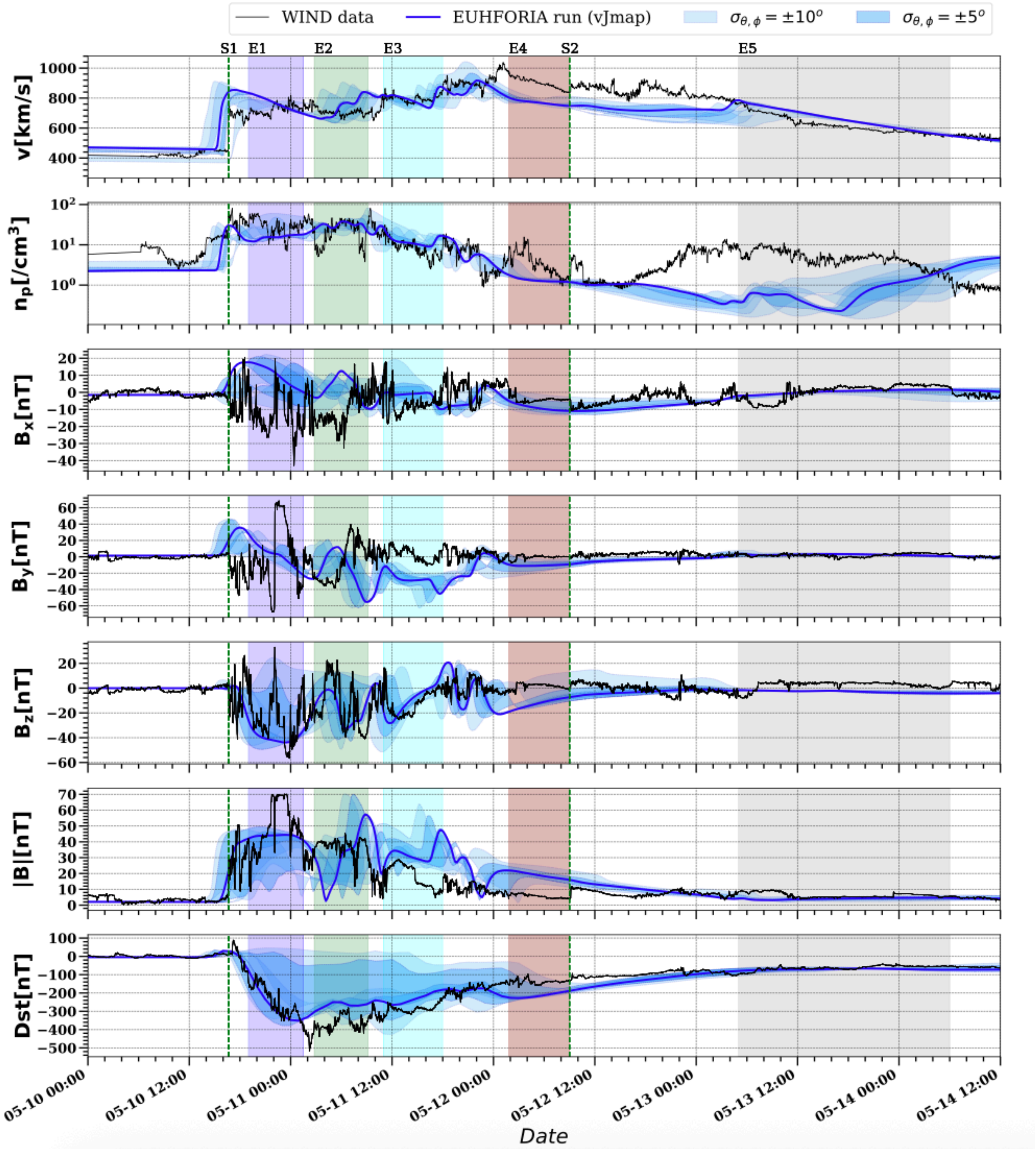

Figure 3. EUHFORIA simulation of the CMEs 1-10 overplotted on the Wind measurements (black) corresponding to the 'vJmap (CME3 GCS)' run. The dashed vertical green lines demarcate the observed shock signatures (S1, S2) and the shaded patches correspond to the observed intervals of the magnetic ejecta (E1, ..., E5) as characterised in Table 3. The shaded blue profile around the solid blue line shows the EUHFORIA predictions for virtual satellites located at 5and 10latitude-longitude shifts relative to Earth. The plot shows top to bottom: the speed v, the proton number density np, the magnetic field components $B_x$, $B_y$ , $B_z$ , the total magnetic field strength |B|, the disturbance storm index Dst, and the plasma beta β.

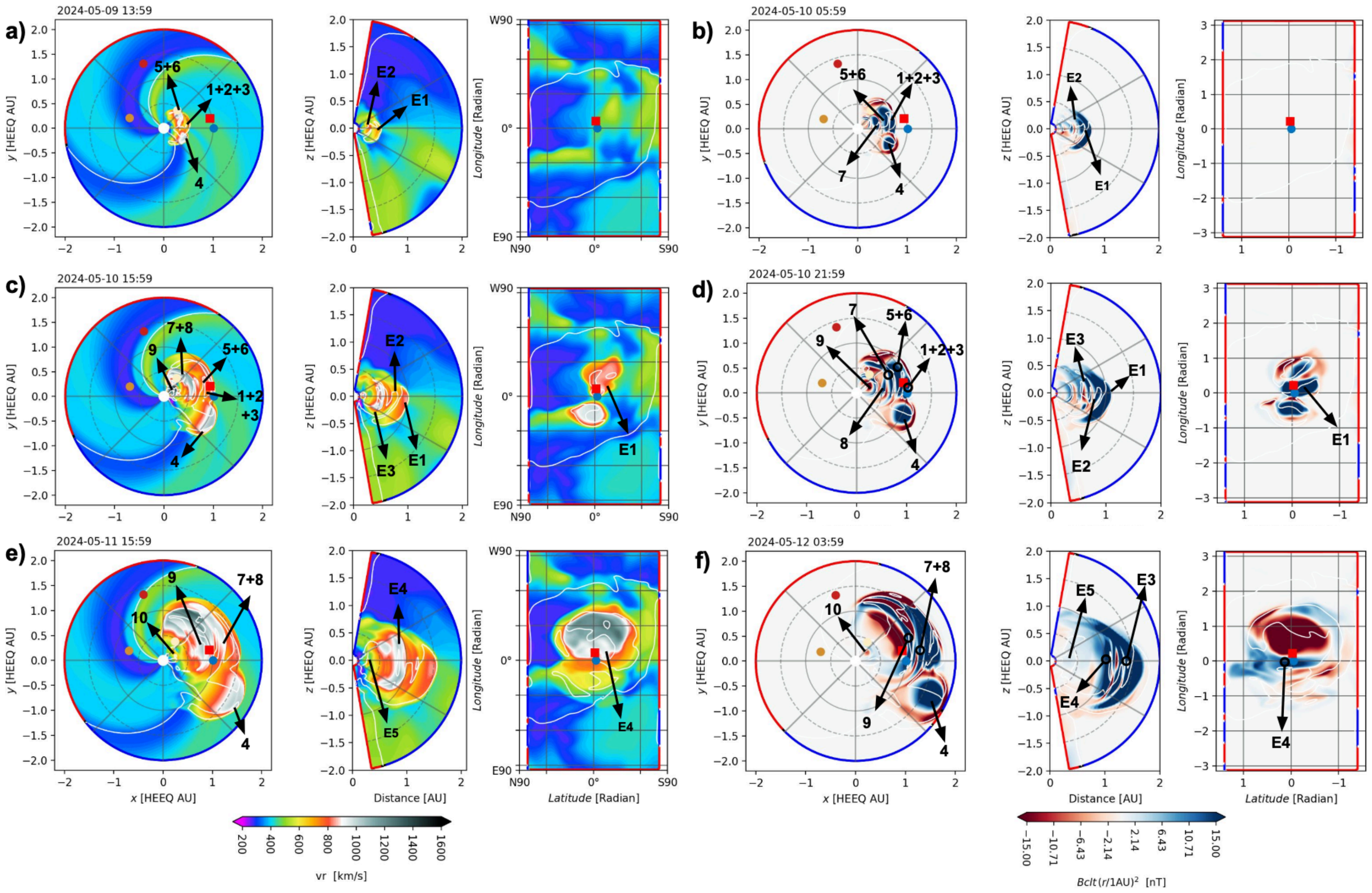

Figure 4. EUHFORI12A simulation, vJmap (CME3 GCS), result of vr (a, c, e) and scaled Bclt, i.e., −Bz , (r/1AU)2 (b, d, f) in the equatorial, meridional, and transversal planes, corresponding to different phases of CME-CME interactions. The subplots correspond to the formation of: (a, b) E1 and E2; (c, d) E3; (e, f) E4 and E5. The blue, orange, and red dots on the equatorial plane denote Earth, Venus and Mercury, respectively. The red square denotes the STEREO-A spacecraft. The animations vr and Bz for the whole period of this superstorm are provided in the supplementary material.

## 5. CONCLUSION AND OUTLOOK

This work attempts to model a super-geomagnetic storm. We adopt a multi-method approach to explain how the CMEs of early May 2024 evolved to create the largest geomagnetic storm in the last two decades. We investigated a series of ten interacting CMEs using a combination of Extreme Ultraviolet (EUV) imaging (solar corona), white-light coronagraph observations (below 0.1 AU) and in-situ solar wind measurements at 1 AU. To better understand these interactions and assess their impact on space weather, we utilised the EUHFORIA model to simulate their propagation and arrival at 1 AU. Our findings highlight the inherent complexity of CME-CME interactions, leading to the non-linear evolution of structures during their propagation through the heliosphere. We characterised all five magnetic ejecta in this interval and proposed a potential mechanism for their formation. We also explained how the relative motion and interaction of the CMEs led to their geomagnetic impact.

We provide a clear list of CMEs involved in this event and correlate them with the identifications reported in previous studies. We add three new CMEs whose crucial contribution to the storm was not accounted for. Between May 8 and 9, six CMEs (CME1 to CME7; except CME4) were launched in quick succession from the AR 13664, which interacted with one another during their transit through the inner heliosphere. CME4 was a filament eruption from AR 11343. This was followed by the launch of another 3 CMEs between late May 9 and May 11: CME8 and 9 from AR 13664 and CME10 from AR 13668. Based on our modelling insights, the following is the interpretation of our results about the CME dynamics:

- CME1 and CME2 merged below 0.1 AU. They started merging with CME3 at ~0.5 AU. This resulted in the formation of E1 with the enhanced v, $B_z$ and |B| during this interval. Although the individual CMEs did not exhibit particularly high velocities near the Sun, their interplanetary interactions significantly enhanced geo-effectiveness, supporting the findings of previous studies such as Soni et al. (2023).

- Although CME3 had a slower speed, it already interacted with the flank of the faster CME4, close to 0.1 AU. This led to partial merging of the CMEs (the flank of CME4 with CME3) and to acceleration of CME3. Even through the limited flank contact, CME4 substantially contributed to: (1) the formation of E1 by increasing the CME3 momentum to merge with CME1 and CME2; and (2) the enhancement of Bz in E1. The presence of CME4 (accelerating CME3) created a low-density region that allowed CME5 to expand more rapidly and begin compressing and merging with E1. This conclusion aligns with the importance of capturing flank encounters, as highlighted in previous studies (Maharana et al. 2022; Palmerio et al. 2023).

• CME5 and CME6 erupted at the same time, and it was difficult to reconstruct them as two separate CMEs in the coronagraph field of view. Injecting them as two separate CMEs facilitated relative motion and interaction, creating E2. The presence of CME6, in addition to CME5, also improved the prediction of Bz in E1, thereby improving Dst by 30%. The trailing part of CME5 compressed by CME7 enhanced the Bz in E2. Enhancement in −Bz resulted from the compression of the following interacting ICMEs, which were previously studied by Pal et al. (2023).

• The interaction of CME8 flank with CME7 enhanced v and Bz in E3 to match the observations.

• The last three CMEs, CME8, CME9 (E4) and CME10 (E5), were launched with much lower speeds compared to CME7 and in the direction further west (more than 25) of the Sun-Earth line. These three CMEs expanded into the low-density regions created by the fast CMEs E3 and E4 in the front, respectively. This resulted in the high-speed, low-magnetic-field regime after May 11 at 18.00. While CME8 and CME9 contributed to enhanced solar wind velocity at 1 AU, their impact on geomagnetic conditions was relatively minor due to a lack of magnetic field rotations in their flank regions, manifesting only as modest Dst dips during the storm's recovery phase.

• Our simulations provide a minimum Dst= −352 nT (reference model estimate is −278 nT; observed is −518 nT), and the magnetic field profile is reproduced well. The above trends in this event closely align with the previous findings by Liu et al. 2019), who emphasised that prolonged eruptive activity from a single region, combined with interplanetary CME-CME interactions, can produce unexpectedly strong magnetic fields even in the absence of418
extreme CME speeds.

Our data-driven EUHFORIA simulations enabled us to estimate the geo-effectiveness (Dst) of the Mother's Day event with up to a 70% accuracy. Notably, extended intervals of southward magnetic field ($B_z < 0$) and elevated solar wind speeds emerged as critical indicators of geomagnetic storm potential, as reflected by significant depressions in the Dst index. These findings emphasise the need to self-consistently simulate the dynamics of CME-CME interactions within the forecasting facility to enhance the accuracy of space-weather predictions. It is difficult to assess the geoeffectiveness of their collective evolution using empirical or analytical methods, as these methods are limited in their ability to capture the dynamics of CME-CME interactions. Hence, MHD modelling improves the predictive accuracy in such cases.

Using the EUHFORIA model, we captured the overall trends in v and $B_z$, two essential ingredients for geoeffectiveness predictions, with high accuracy. However, achieving a one-to-one match with in-situ observations remains challenging due to multiple factors that complicate the full replication of the observed solar wind signatures and CME dynamics at 1 AU. The following could be some of the limitations:

• The level of uncertainty associated with the CME input parameters is the major bottleneck. The complex nature of CME-CME interactions depends on the relative positions and speeds of individual CMEs. As the CMEs in this event interval were either halo or partial-halo, the uncertainty associated with their locations, geometric parameters, and speed estimates is a major determinant of prediction accuracy.

• The spheromak CME model used in this study, although a suitable model to quickly constrain in the operational setup, does not fully represent the CME morphology and the internal magnetic structure. Hence, our model results do not fully match the Bx and By components, nor the total magnetic field |B|. The flux rope orientation during its heliospheric propagation could also be altered if the mass/momentum is not accurately constrained (Asvestari et al., 2022).

• In our modelling, we have used the steady-state solar wind; however, in reality, the wind is dynamically evolving. Hence, the interaction of the CMEs with the dynamic solar wind could result in more temporally varying features, which we cannot capture in our current setup (Samara et al., 2025).

• The MHD framework of EUHFORIA restricts the modelling of mesoscale processes like magnetic reconnection and other processes (Kilpua et al., 2017), which form the fluctuating plasma and magnetic field features in the sheaths. The observations show that the intervals between E3 and E4 and between E4 and E5 are sheath regions driven by CME8 and CME9, respectively. However, we were unable to accurately capture the magnetic field and plasma properties during that interval through our MHD modelling. In addition, there are multiple intervals of coherent magnetic field structures emerging from fluctuations within E1, E2, and E3 characterised by Pal et al. (2025)- modelling them is beyond the MHD framework.

• Resolution of simulation plays a crucial role in the accuracy of predictions. In the EUHFORIA model, increasing the radial resolution from 256 to 512 cells between 0.1 and 2 AU and the angular resolution from 4 to 2 enhanced feature resolution and captured more dynamical detail. Resolving the features improved the accuracy of Bz prediction, which was underestimated otherwise; however, the speed of arrival (S1) and other magnetic field components are slightly overestimated (see Fig. A4).

### 5.1. Outlook

It goes without saying, however, this must be re-emphasised in the context of this study: In post-event analysis, we have the luxury of performing the CME reconstruction multiple times to improve parameter estimation and modelling of superstorms- a liberty that real-time forecasters do not have. This calls for improved methodologies to add uncertainty bounds on the CME parameters in catalogues- thereby informing a sharper range for ensemble modelling.

The scenario becomes increasingly complex when events involve multiple CME-CME interactions; forecasters can’t make an informed estimate of which parameters to adjust. Ensemble modelling (e.g., Flossie et al., 2025; Singh et al., 2020) can become cumbersome while dealing with ten CMEs. Hence, the potential for employing machine learning algorithms to learn the natural dynamics between the parameters of two CMEs must be explored to improve predictions. We foresee improving the modelling of, especially, flank CMEs with advanced CME models, e.g., the FRi3D model using a recently developed, more stable numerical methodology (see e.g., Flossie et al. 2025).

ACKNOWLEDGEMENTS

AM and S.Poedts are funded by the European Union. However, the views and opinions expressed are those of the author(s) only and do not necessarily reflect those of the European Union or ERCEA. Neither the European Union nor the granting authority can be held responsible. His project (Open SESAME) has received funding under the Horizon Europe programme (ERC-AdG agreement No 101141362). These results were also obtained in the framework of the projects C16/24/010 (C1 project Internal Funds KU Leuven), G0B5823N and G002523N (WEAVE) (FWO-Vlaanderen), and 4000145223 (SIDC Data Exploitation (SIDEX2), ESA Prodex). We used the VSC- Flemish Supercomputer Center infrastructure for the computations, funded by the Hercules Foundation and the Flemish Government, department EWI.

APPENDIX

The GCS reconstructions are provided in Fig. A1.

The post-eruption AR for the CMEs is shown in Fig. A2.

The EUHFORIA simulations of sequential addition of CMEs are provided in Fig. A3.

The EUHFORIA simulation results comparing the resolutions- medium (512 grid cells in radial and 2angular) and low (256 grid cells in radial and 4angular)- are shown in Fig. A4.

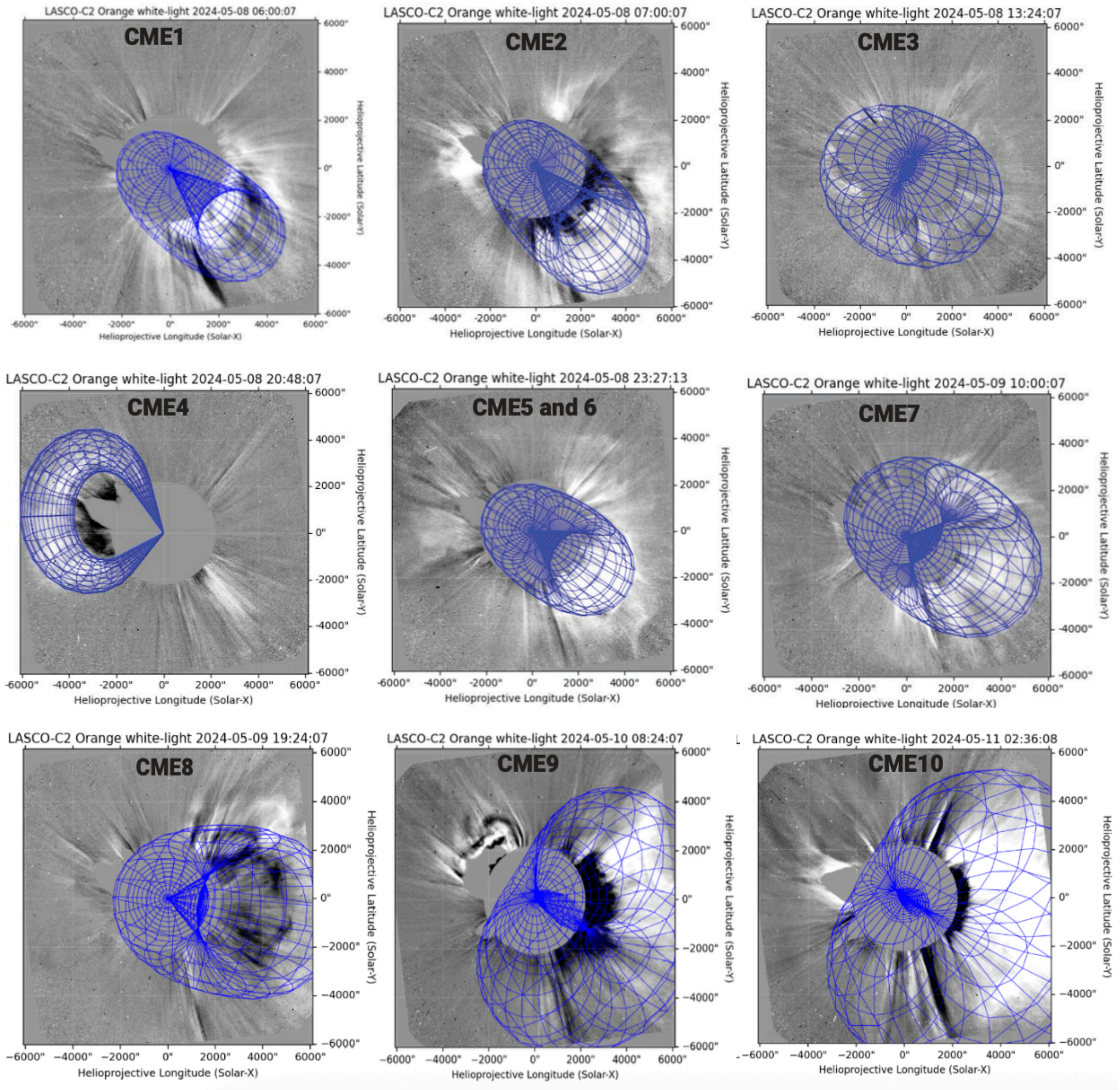

Figure A1. Graduated Cylindrical Shell (GCS) forward-model reconstructions of the ten successive coronal mass ejections (CME1 to CME9) associated with the eruptive activity of NOAA AR 13664 and 13668 (CME9) during 8–10 May 2024, overlaid on SOHO/LASCO-C2 orange white-light coronagraph images. Each panel shows the best-fit 3D flux-rope geometry

(blue wireframe) projected onto the plane of the sky at the time indicated in the panel header. The GCS model constrains the CME apex height, angular width, tilt, and propagation direction by fitting the leading edge and cavity structure visible in the coronagraph data. The sequence reveals notable event-to-event variability in CME geometry, including changes in axis orientation, expansion asymmetry, and halo extent, reflecting the progressive restructuring of the source region and the complex 3D evolution of the eruptive flux systems.

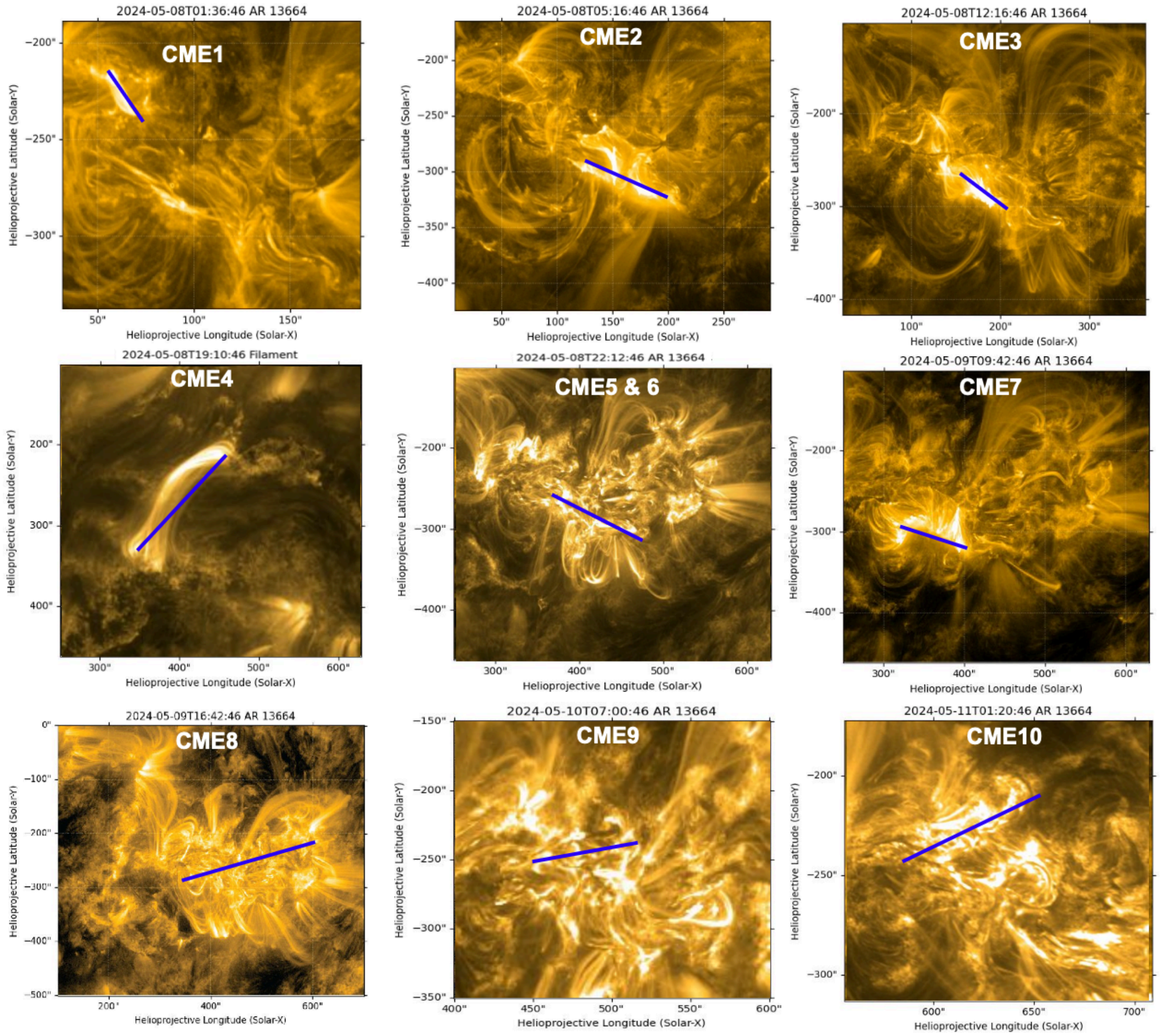

Figure A2. Multi-epoch SDO/AIA 171 ˚A observations illustrating the sequence of ten successive coronal mass ejection (CME) source-region eruptions (CME1 to CME9) from NOAA Active Region AR 13664 and AR 13668 during 8–11 May 2024. Each panel shows the low-coronal morphology at the onset of eruption, highlighting the evolving magnetic architecture, including bright post-flare arcades, twisted coronal loops, and filament-channel structures associated with each CME. The timestamps at the top of each panel indicate the observation time (UT) at the initiation of the respective event. The blue solid lines mark the inferred primary

orientation or propagation axis of the erupting magnetic structure, serving as a proxy for the projected direction and tilt of the pre-eruptive flux system. The sequence reveals progressive structural complexity and repeated destabilisation of the same active region, suggesting the buildup and release of magnetic energy through successive reconnection episodes.

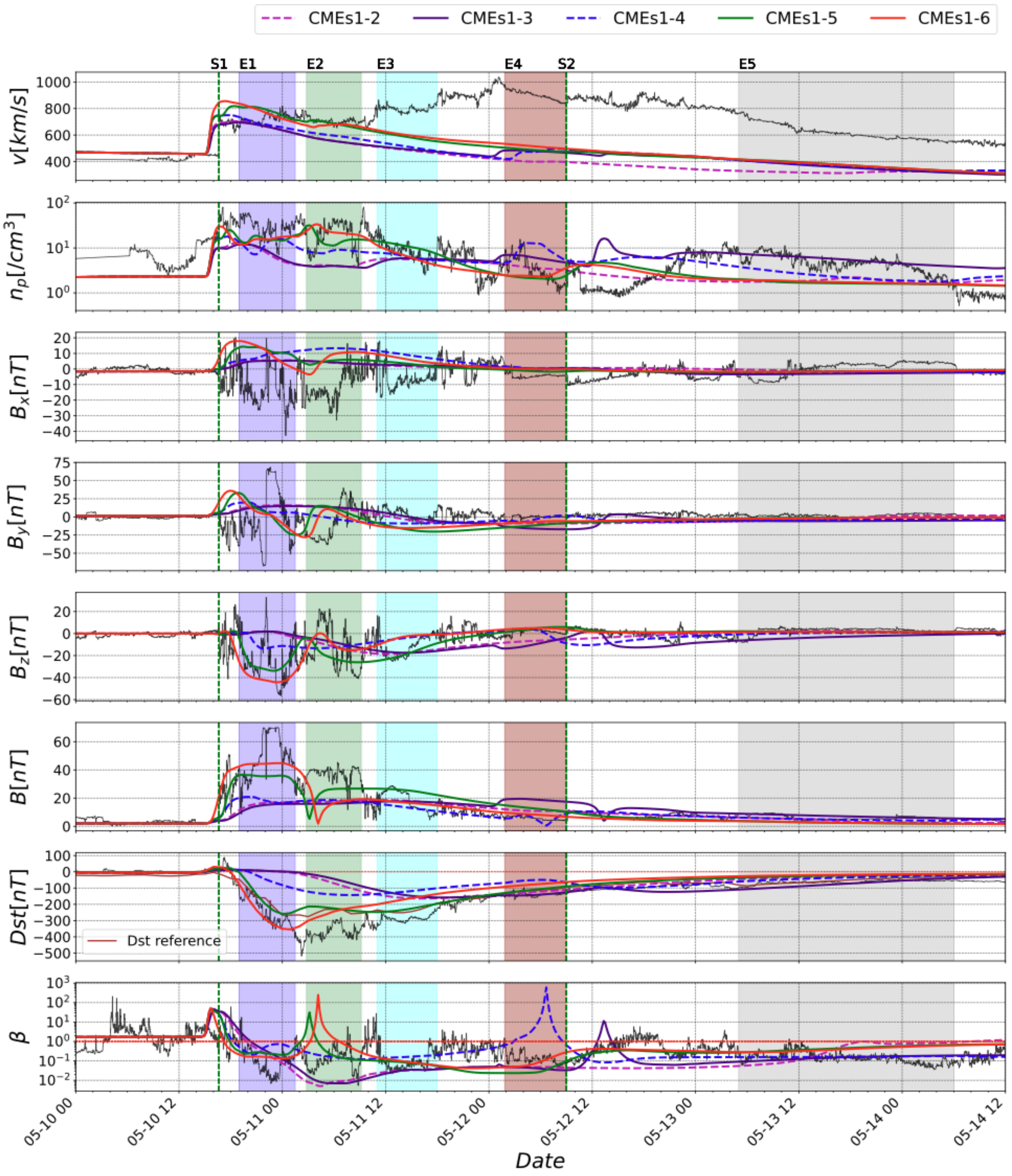

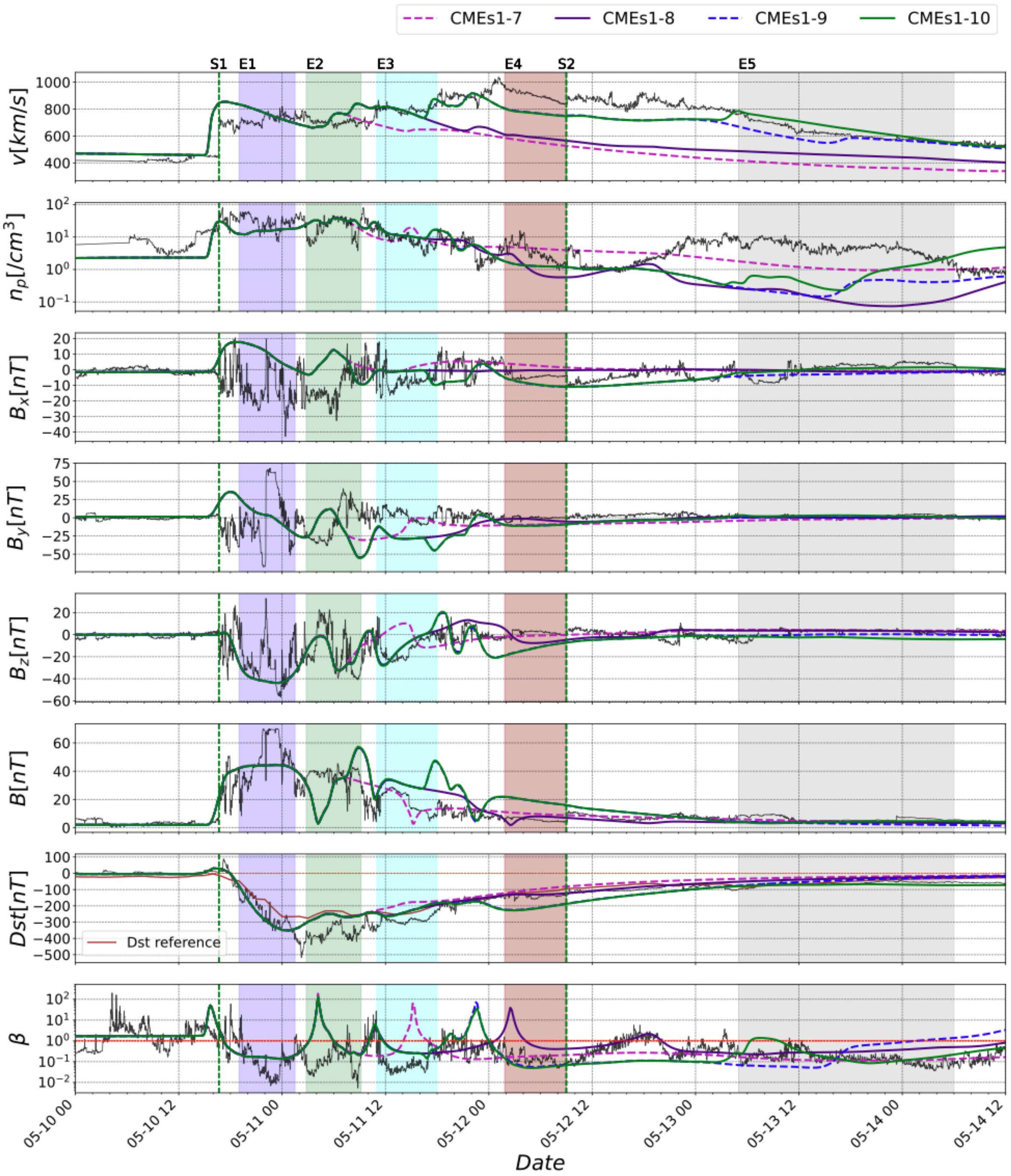

Figure A3. EUHFORIA simulation of the CMEs 1-3, 1-4, 1-5, and 1-6 (top) & CMEs 1-7, 1-8, and 1-9 (bottom) overplotted on the Wind measurements (black). The dashed vertical green lines demarcate the observed shock signatures (S1, S2) and the shaded patches correspond to the observed intervals of the magnetic ejecta (E1, ..., E5) as characterised in Table 3. The plot shows top to bottom: the speed v, the proton number density np, the magnetic field components Bx, By , Bz , the total magnetic field strength |B|, the disturbance storm index Dst, and the plasma beta β.

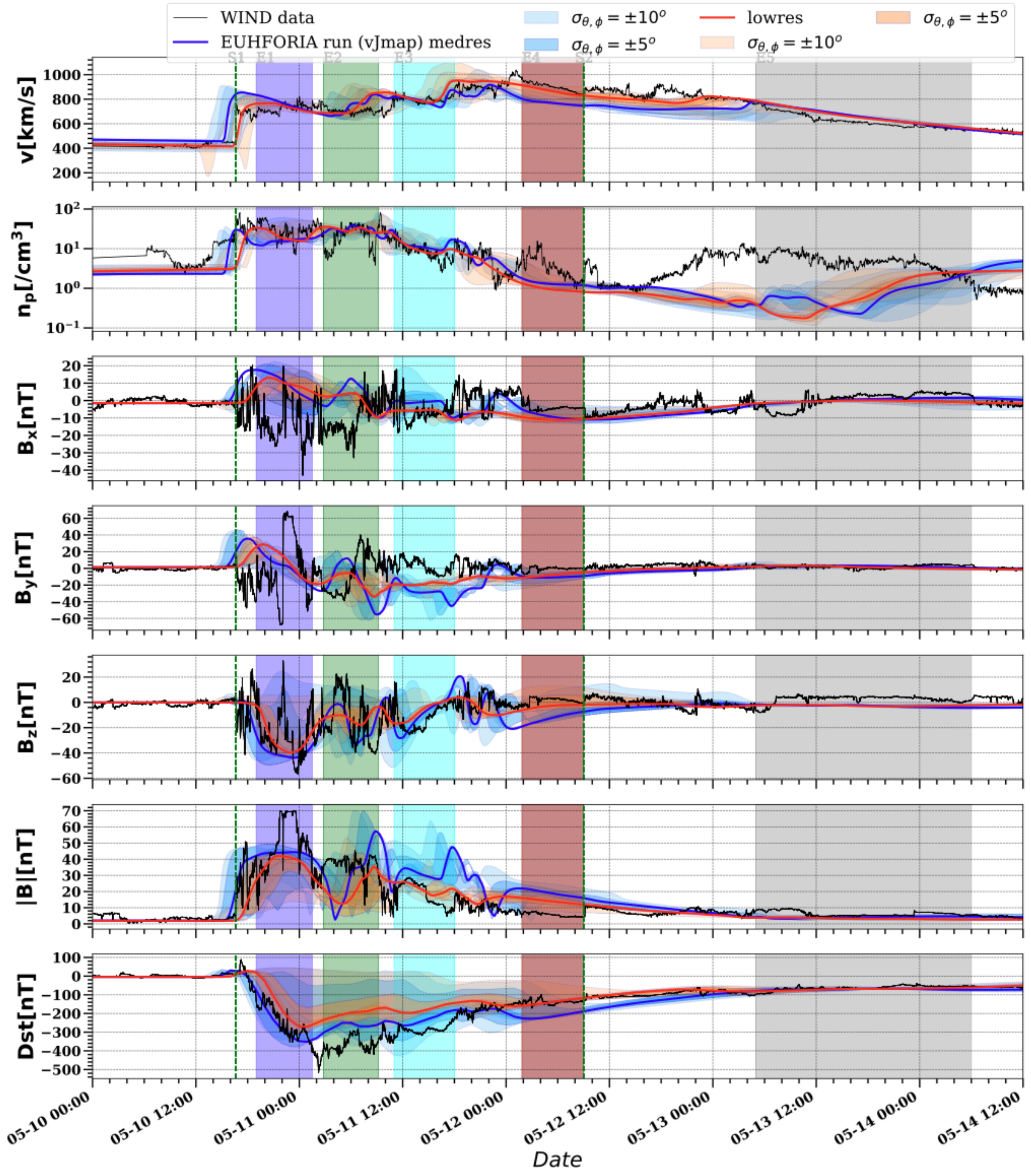

Figure A4. The EUHFORIA simulations with all ten CMEs in medium and low resolution. The plot description is similar to Fig. 3.